\documentclass[final,5p,times,twocolumn]{elsarticle}

\usepackage{amssymb}
\usepackage{amsmath}
\usepackage[utf8]{inputenc}
\usepackage[T1]{fontenc}

 \usepackage{lineno}

\usepackage{hyperref}
\hypersetup{
    colorlinks=true,
    linkcolor=blue}

\usepackage{array,tabularx,caption}    
\journal{NIM A}

\begin{document}
\begin{frontmatter}



\title{Photocathode characterisation for robust PICOSEC Micromegas precise-timing  detectors}

\author[inst1,inst2]{M. Lisowska\corref{cor1}}
\cortext[cor1]{Corresponding author.}
\ead{marta.lisowska@cern.ch}
\author[inst3]{R. Aleksan}
\author[inst4,inst5]{Y. Angelis}
\author[inst3]{S. Aune}
\author[inst6]{J. Bortfeldt}
\author[inst1]{F. Brunbauer}
\author[inst7,inst8]{M. Brunoldi}
\author[inst4,inst5]{E. Chatzianagnostou}
\author[inst9]{J. Datta}
\author[inst10]{K. Dehmelt}
\author[inst11]{G. Fanourakis}
\author[inst1]{S. Ferry}
\author[inst7,inst8]{D. Fiorina\fnref{fn1}}
\author[inst1,inst12]{K. J. Floethner}
\author[inst13]{M. Gallinaro}
\author[inst14]{F. Garcia}
\author[inst3]{I. Giomataris}
\author[inst10]{K. Gnanvo}
\author[inst3]{F.J. Iguaz\fnref{fn2}}
\author[inst1]{D. Janssens}
\author[inst3]{A. Kallitsopoulou}
\author[inst16]{M. Kovacic}
\author[inst10]{B. Kross}
\author[inst17]{C.C. Lai}
\author[inst3]{P. Legou}
\author[inst18]{J. Liu}
\author[inst12,inst19]{M. Lupberger}
\author[inst1,inst4]{I. Maniatis\fnref{fn3}}
\author[inst10]{J. McKisson}
\author[inst18]{Y. Meng}
\author[inst1,inst19]{H. Muller}
\author[inst1]{R. De Oliveira}
\author[inst1]{E. Oliveri}
\author[inst1,inst20]{G. Orlandini}
\author[inst10]{A. Pandey}
\author[inst3]{T. Papaevangelou}
\author[inst21]{M. Pomorski}
\author[inst1,inst22]{M. Robert}
\author[inst1]{L. Ropelewski}
\author[inst4,inst5]{D. Sampsonidis} 
\author[inst1]{L. Scharenberg}
\author[inst1]{T. Schneider}
\author[inst21]{E. Scorsone}
\author[inst1,inst3]{L. Sohl\fnref{fn4}}
\author[inst1]{M. van Stenis}
\author[inst23]{Y. Tsipolitis}
\author[inst4,inst5]{S. Tzamarias}
\author[inst24]{A. Utrobicic}
\author[inst7,inst8]{I. Vai}
\author[inst1]{R. Veenhof}
\author[inst1]{L. Viezzi}
\author[inst7,inst8]{P. Vitulo}
\author[inst1,inst25]{C. Volpato}
\author[inst18]{X. Wang}
\author[inst1,inst26]{S. White}
\author[inst10]{W. Xi}
\author[inst18]{Z. Zhang}
\author[inst18]{Y. Zhou}

\affiliation[inst1]{organization={European Organization for Nuclear Research (CERN), 1211 Geneve 23, Switzerland}}

\affiliation[inst2]{organization={Université Paris-Saclay, F-91191 Gif-sur-Yvette, France}}

\affiliation[inst3]{organization={IRFU, CEA, Université Paris-Saclay, F-91191 Gif-sur-Yvette, France}}

\affiliation[inst4]{organization={Department of Physics, Aristotle University of Thessaloniki, University Campus, GR-54124, Thessaloniki, Greece}}

\affiliation[inst5]{organization={Center for Interdisciplinary Research and Innovation (CIRI-AUTH), Thessaloniki 57001, Greece}}

\affiliation[inst6]{organization={Department for Medical Physics, Ludwig Maximilian University of Munich, Am Coulombwall 1, 85748 Garching, Germany}}

\affiliation[inst7]{organization={Dipartimento di Fisica, Università di Pavia, Via Bassi 6, 27100 Pavia, Italy}}

\affiliation[inst8]{organization={INFN Sezione di Pavia, Via Bassi 6, 27100 Pavia, Italy}}

\affiliation[inst9]{organization={Department of Physics and Astronomy, Stony Brook University, Stony Brook, NY 11794-3800, USA}}

\affiliation[inst10]{organization={Jefferson Lab, 12000 Jefferson Avenue, Newport News, VA 23606, USA}}

\affiliation[inst11]{organization={Institute of Nuclear and Particle Physics, NCSR Demokritos, GR-15341 Agia Paraskevi, Attiki, Greece}}

\affiliation[inst12]{organization={Helmholtz-Institut für Strahlen- und Kernphysik, University of Bonn, Nußallee 14–16, 53115 Bonn, Germany}}

\affiliation[inst13]{organization={Laboratório de Instrumentacão e Física Experimental de Partículas, Lisbon, Portugal}}

\affiliation[inst14]{organization={Helsinki Institute of Physics, University of Helsinki, FI-00014 Helsinki, Finland}}

\affiliation[inst16]{organization={University of Zagreb, Faculty of Electrical Engineering and Computing, 10000 Zagreb, Croatia}}

\affiliation[inst17]{organization={European Spallation Source (ESS), Partikelgatan 2, 224 84 Lund, Sweden}}

\affiliation[inst18]{organization={State Key Laboratory of Particle Detection and Electronics, University of Science and Technology of China, Hefei 230026, China}}

\affiliation[inst19]{organization={Physikalisches Institut, University of Bonn, Nußallee 12, 53115 Bonn, Germany}}

\affiliation[inst20]{organization={Friedrich-Alexander-Universität Erlangen-Nürnberg, Schloßplatz 4, 91054 Erlangen, Germany}}

\affiliation[inst21]{organization={CEA-LIST, Diamond Sensors Laboratory, CEA Saclay, F-91191 Gif-sur-Yvette, France}}

\affiliation[inst22]{organization={Queen’s University, Kingston, Ontario, Canada}}

\affiliation[inst23]{organization={National Technical University of Athens, Athens, Greece}}

\affiliation[inst24]{organization={Ruđer Bošković Institute, Bijenička cesta 54., 10 000 Zagreb, Croatia}}

\affiliation[inst25]{organization={Department of Physics and Astronomy, University of Florence, Via Giovanni Sansone 1, 50019 Sesto Fiorentino, Italy}}

\affiliation[inst26]{organization={University of Virginia, USA}}

\fntext[fn1]{Now at Gran Sasso Science Institute, Viale F. Crispi, 7 67100 L'Aquila, Italy.}
\fntext[fn2]{Now at SOLEIL Synchrotron, L’Orme des Merisiers, Départementale 128, 91190 Saint-Aubin, France.}
\fntext[fn3]{Now at Department of Particle Physics and Astronomy, Weizmann Institute of Science, Hrzl st. 234, Rehovot, 7610001, Israel.}
\fntext[fn4]{Now at TÜV NORD EnSys GmbH \& Co. KG.}

\begin{abstract}

The PICOSEC Micromegas detector is a~precise-timing gaseous detector based on a~Cherenkov radiator coupled with a~semi-transparent photocathode and a~Micromegas amplifying structure, targeting a~time resolution of tens of picoseconds for minimum ionising particles.
Initial single-pad prototypes have demonstrated a~time resolution below $\sigma$ = 25 ps, prompting ongoing developments to adapt the concept for High Energy Physics applications, where sub-nanosecond precision is essential for event separation, improved track reconstruction and particle identification.
The achieved performance is being transferred to robust multi-channel detector modules suitable for large-area detection systems requiring excellent timing precision.
To enhance the robustness and stability of the PICOSEC Micromegas detector, research on robust carbon-based photocathodes, including Diamond-Like Carbon (DLC) and Boron Carbide (B$_4$C), is pursued.
Results from prototypes equipped with DLC and B$_4$C photocathodes exhibited a~time resolution of  $\sigma$~$\approx$~32 ps and  $\sigma$ $\approx$ 34.5 ps, respectively.
Efforts dedicated to improve detector robustness and stability enhance the feasibility of the PICOSEC Micromegas concept for large experiments, ensuring sustained performance while maintaining excellent timing precision.

\end{abstract}




\end{frontmatter}





\newpage
~
\newpage

\section{Introduction}
\label{sec:sample1}

The intense interest in advancing technologies for precise-timing detectors has been driven by demanding environments anticipated in forthcoming High Energy Physics experiments~\cite{ECFA}.
Sub-nanosecond precision is crucial for distinguishing events occurring in rapid succession, improving track reconstruction and enabling particle identification through Time of Flight measurements.
Meeting the criteria of achieving an excellent time resolution, ensuring stable long-term operation and providing large-area coverage is crucial for adapting the concept to be used in future experiments.
The PICOSEC Micromegas (hereafter PICOSEC) project \cite{firstPicosecPaper} undertakes the efforts to develop a~robust multi-channel gaseous detector, targeting a~time resolution of $\mathcal{O}(10)$ ps for Minimum Ionising Particles (MIPs).
Initial single-pad prototypes equipped with Cesium Iodide (CsI) photocathodes have demonstrated a~time resolution below  $\sigma$ = 25 ps \cite{firstPicosecPaper}, prompting further developments such as detector design optimisation \cite{AntonijaSinglePad}, enhanced stability and robustness \cite{LukasPhDThesis,myMScThesis,myMPGD,XuPhotocathodes}, as well as scaling to larger areas \cite{LukasPhDThesis,multipadPicosecPaper,NDIP,AntonijaMPGD}.
While the CsI photocathode exhibits high quantum efficiency (QE) and ultraviolet (UV) sensitivity, it is prone to damage from ion backflow and discharges, as well as humidity.
Although preliminary measurements of alternative approaches have been conducted previously \cite{LukasPhDThesis,myMScThesis,myMPGD,XuPhotocathodes}, a~comprehensive characterisation of different carbon-based photocathodes has not been reported.
This work aims to advance the development of robust photocathodes for PICOSEC detectors while maintaining an excellent time resolution.

Within the scope of this paper, the characterisation of three different photocathode materials - CsI, Diamond-Like Carbon (DLC) and Boron Carbide (B$_4$C) - and their performances as candidates for the PICOSEC detectors are reported.
The samples were manufactured by various research institutes, including the European Organization for Nuclear Research (CERN), the University of Science and Technology of China (USTC), the French Atomic Energy Commission (CEA) and the European Spallation Source (ESS).

\section{PICOSEC Micromegas detection concept}
\label{sec:2}

\begin{figure*}[!t]
\begin{center}
\includegraphics[width=\textwidth]{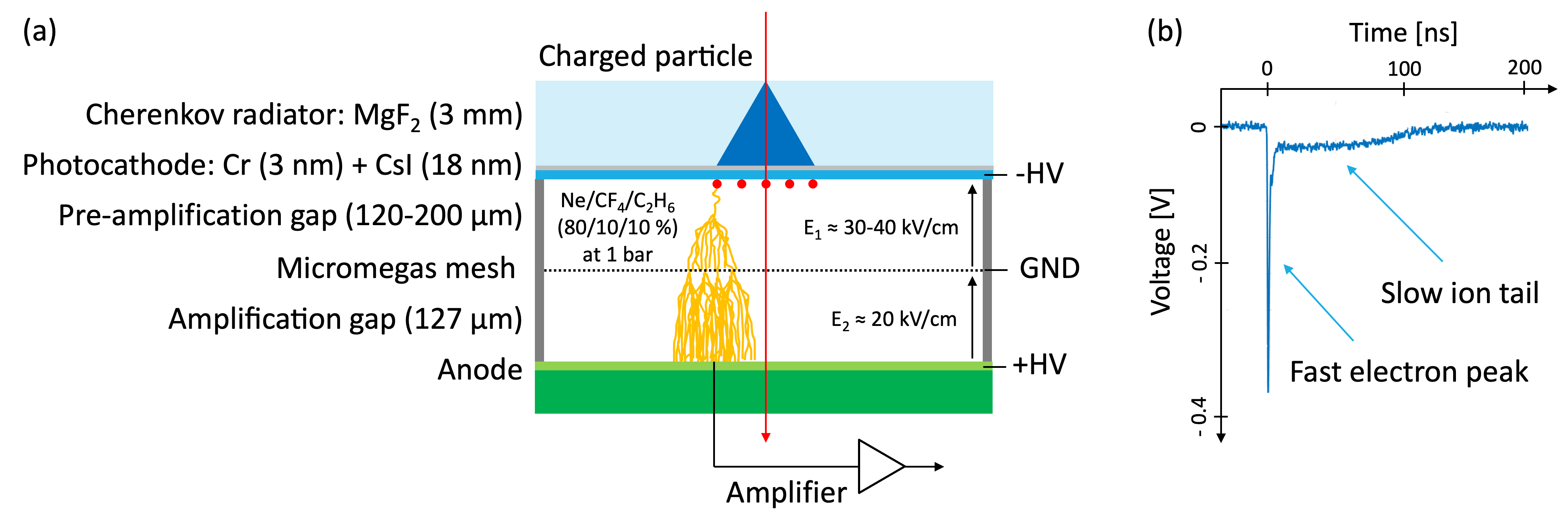}
\end{center}
\caption{(a) PICOSEC detection concept: A~charged particle passing through a~Cherenkov radiator generates UV photons, which are converted into electrons on a~photocathode. The electrons are multiplied in the gas volume in two stages and the movement of charge in the amplification gap induces a~signal on the anode.  Note that the figure are not drawn to scale. (b) Typical PICOSEC waveform: The signal displays a~fast electron peak and a~slow ion tail.}
\label{PicosecDetectionConcept}
\end{figure*}

The PICOSEC Micromegas detection concept is detailed in~\cite{firstPicosecPaper} and depicted in Fig.~\ref{PicosecDetectionConcept}a.
The detector is designed to minimise time jitter from primary ionisation in the gas by using a~Cherenkov radiator.
A charged particle passing through the radiator generates a~cone of UV photons which are partially converted into  primary electrons on a~photocathode coated on the crystal.
Typically, the radiator in a~PICOSEC detector is a~3 mm thick Magnesium Fluoride (MgF$_2$) crystal which transmits light above 120 nm.
The window is coated with a~3 nm thick chromium (Cr) conductive interfacial layer that serves as a~contact for the high voltage (HV).
The UV photon-to-electron converter is an 18 nm thick semi-transparent CsI photocathode known for its high QE for photons below 200 nm \cite{firstPicosecPaper}.
All primary electrons are created at the same surface, eliminating uncertainty about the ionisation location and thereby minimising time jitter.

The gas volume is filled with a~mixture optimised for good time resolution by the COMPASS experiment  \cite{COMPASS}, consisting of 80\% Ne, along with 10\% C$_2$H$_6$ and 10\% CF$_4$ to reduce electron diffusion, at ambient pressure.
A~stainless steel calendared woven mesh with 18~µm wire diameter, 45~µm opening, 30~µm thickness and about 50\% optical transparency \cite{firstPicosecPaper} serves a dual purpose: it protects the photocathode from ion feedback that could degrade its performance and ensures stability by preventing discharges.
The pre-amplification gap between the cathode and the mesh is defined by a spacer made of two-layer copper-clad polyimide.
The PICOSEC detector operates at HV levels close to, but not exceeding, the discharge limit.
The typical electric fields across the pre-amplification and amplification gaps are around 30-40~kV/cm and 20~kV/cm, correspondingly.
In the presence of a~strong electric field, the extracted electrons successively ionise gas molecules, causing electron multiplication, which first occurs in the pre-amplification gap and, after passing through the mesh, continues in the amplification gap.
The total gain achieved is on the level of $\mathcal{O}(10^{5} - 10^{6})$.

The amplified electrons move towards the anode, while the ions travel to the mesh.
Their movement induces a~signal on the anode, which is then amplified and read out by a~digitiser.
A typical PICOSEC waveform displaying a~fast electron peak and a~slow ion tail is illustrated in Fig.~\ref{PicosecDetectionConcept}b.
The leading edge of the electron peak determines a~Signal Arrival Time (SAT).

\begin{figure}[!t]
\begin{center}
\includegraphics[width=8cm]{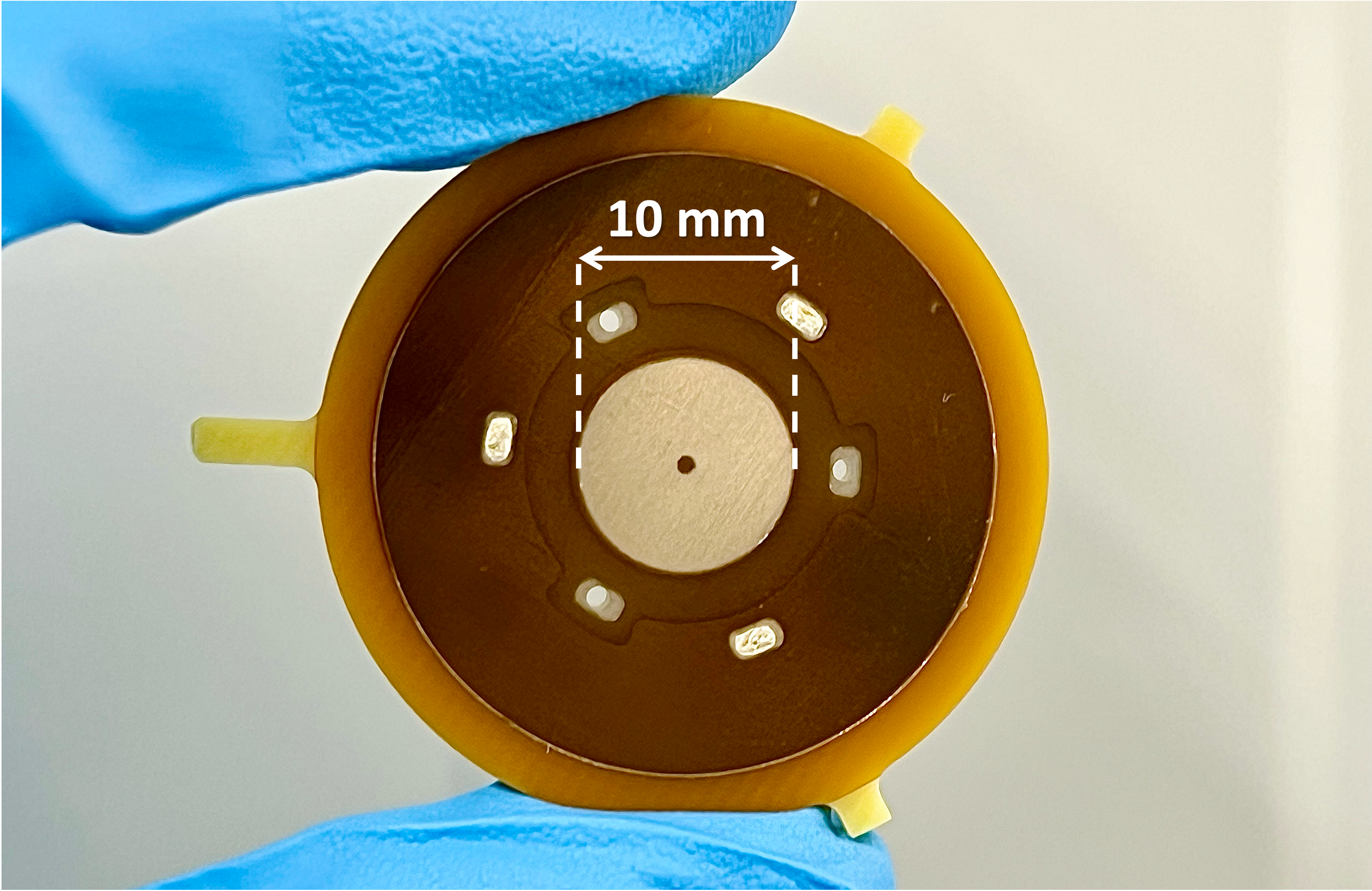}
\end{center}
\caption{Single-channel detector board with a~10 mm diameter active area. The Micromegas board consists of an FR4 printed circuit board (PCB) with two copper layers, featuring a copper readout pad and a ground (GND) plane ring on the top side. A mesh is sandwiched between two polyimide coverlays, both of which have openings for the active area, gas circulation, and mesh-GND connection. A pillar in the centre of the pad supports the mesh.}
\label{singlePadPhoto}
\end{figure}

Several optimisation studies were conducted on the single-pad PICOSEC detector to enhance HV stability, reduce noise level, improve signal integrity and achieve uniform timing response, using a~simplified assembly process.
The  design includes improvements to the detector board, vessel, mechanical parts and electrical connections for both HV and signal.
A~new single-pad detector board with a~10 mm diameter active area is shown in Fig.~\ref{singlePadPhoto}.
A comprehensive overview of the mechanical design can be found in reference \cite{AntonijaSinglePad}.

To enhance the robustness and stability of the PICOSEC detector, research on robust photocathodes is underway.
The initial prototype involved a~CsI photocathode due to its high QE and UV sensitivity, nonetheless, the material is vulnerable to damage from ion blackflow, discharges and humidity.
The robustness of the photocathode is crucial for maintaining detector efficiency and timing resolution during long-term operation.
Potential alternatives include using protective layers or other materials, with carbon-based photocathodes like DLC and B$_4$C being the most promising candidates.

\section{Experimental methodology}
\label{sec:3}

R\&D activities within the PICOSEC project have covered aspects ranging from design and production, through assembly, to measurements in laboratory conditions as well as with particle beams.
Prototypes were assembled in a~clean-room environment and tested in the laboratory.
The ASSET (A~Small Sample Evaporation Test) setup was developed to characterise photocathodes, facilitating transparency and QE measurements, as well as ageing studies \cite{myMScThesis,ALICE_RICH,ALICE_CsI}.
The time resolution of the detectors was evaluated using 150~GeV/c muon beams at the CERN SPS H4 beamline.

\subsection{ASSET photocathode characterisation device}

\begin{figure*}[!t]
\begin{center}
\includegraphics[width=\textwidth]{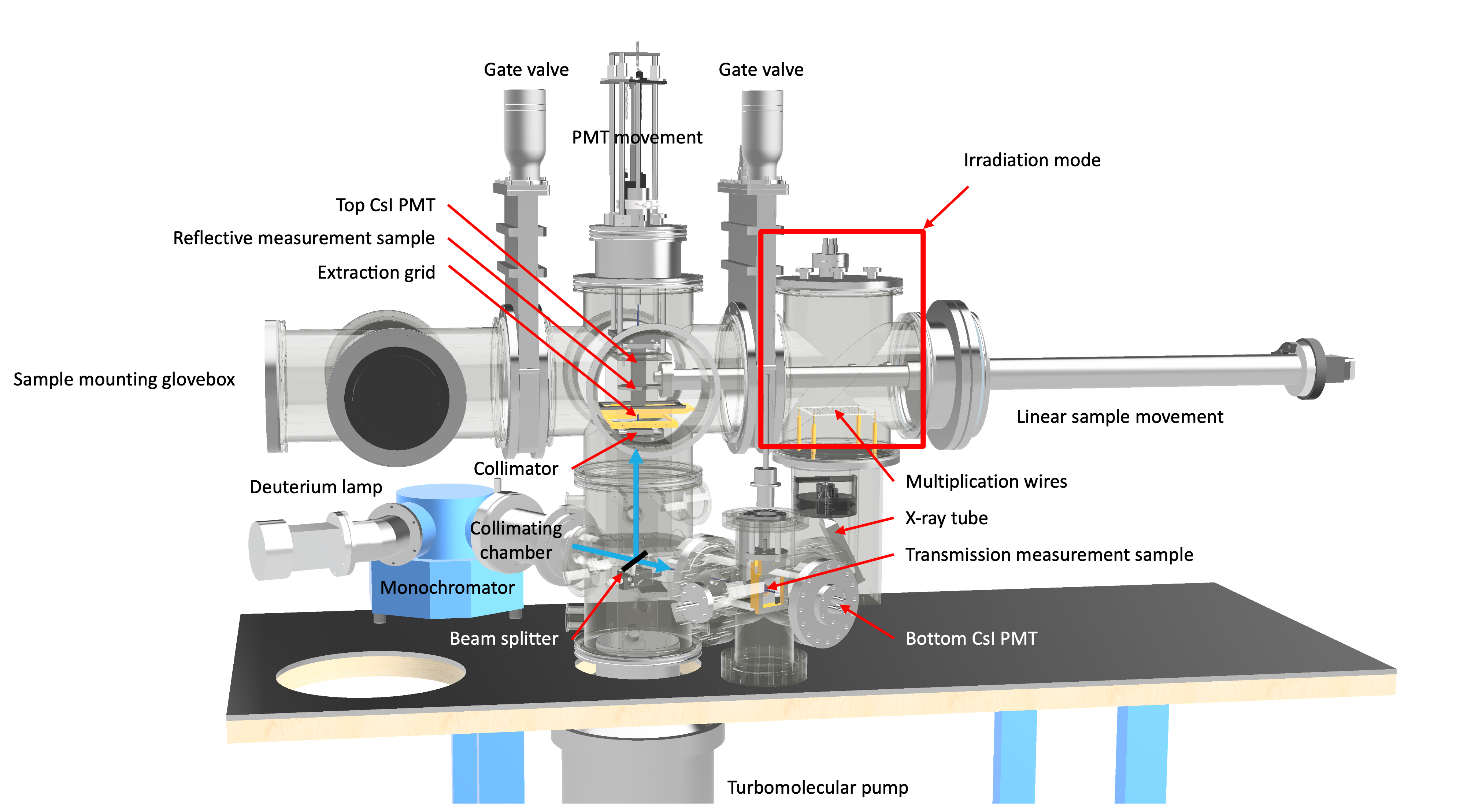}
\end{center}
\caption{Overview of the ASSET photocathode characterisation device: The setup uses UV light from a~deuterium lamp and a~VUV monochromator, features a~collimating mirror chamber, a~beam splitter and two calibrated CsI PMTs to measure light intensity and stability within a~wavelength range of 120-200 nm, peaking at 160-180 nm. It allows for measurements in reflective and transmission mode, as well as studies of the ageing process due to ion backflow \cite{myMScThesis}.}
\label{ASSET}
\end{figure*}

The ASSET setup was developed to characterise photocathodes for gaseous radiation detectors, as in the PICOSEC project~\cite{myMScThesis}.
The primary aims are to quantify QE and transparency of the photocathodes as well as their degradation due to ion backflow.
Figure~\ref{ASSET} provides an overview of the ASSET setup.

Measurements utilise UV light from a~system consisting of a~deuterium lamp (McPherson, Model 632 \cite{deuteriumLamp}) and a~Vacuum UltraViolet (VUV) monochromator (McPherson, Monarch \cite{Monochromator}).
The system is flushed with high-purity nitrogen to prevent absorption of the short-wavelength portion of the UV spectrum. 
The monochromator is attached to the measurement chamber through an extension containing a~collimating mirror chamber that focuses the light.
A beam splitter divides the light into two beams. 
Two calibrated CsI photomultiplier tubes (PMTs) register the light intensity: one measures the amount of light at the sample position, while the other monitors light stability during measurements.
The wavelength range for measurements is limited to about 120 nm on the low-wavelength side by the transparency of the MgF$_2$ lens and to about 200 nm on the high-wavelength side by the deuterium lamp's light intensity.
The highest light intensity is observed at 160-180 nm.

The ASSET setup allows for measurements in reflective and transmission modes.
The measurements are conducted in a~vacuum of 10$^{-6}$~mbar, achieved using pre-pumps and turbomolecular pumps. 
In both modes, the extraction grids record the current generated by electrons extracted from the photocathodes.
The reflective mode measures the QE of photocathodes, while the transmission mode evaluates both QE and transparency.

The main function of the ASSET setup is the measurement of current from photocathode samples.
The QE is calculated from current measurements using the formula:
\begin{equation}
QE = \frac{N_\text{e}}{ N_{\text{ph}}} = \frac{\frac{ I_{\text{s}}}{e}}{\frac{I_{\text{PMT}}}{e~\cdot~C_{\text{PMT}}}} = \frac{ I_{\text{s}} \cdot C_{\text{PMT}}}{I_{\text{PMT}}},
\end{equation}
where
\( N_\text{e} \) is the number of electrons extracted from the sample (measured on the sample),
 \( N_{\text{ph}} \)~is the number of photons that arrive at the sample (measured on the PMT),
 \( I_{\text{s}} \) is the current flowing from the sample to the wire grid, measured by a~Keithley 6487 picoammeter~\cite{Keithley},
  \( I_{\text{PMT}} \) is the current measured on the PMT using the picoammeter,
 \( C_{\text{PMT}} \) is the wavelength-dependent calibration factor for the PMT, as provided in the datasheet of the device and
 \( e  \) is the elementary charge.

The current measured by the PMT is used to calculate the transparency $T$ of the samples, given by: 
\begin{equation}
T=\frac{ {I}_{\text{PMT,~s.~in}} } { {I}_{\text{PMT,~s.~out}} },
\end{equation}
where
 ${I}_{\text{PMT,~s.~in}}$ is the PMT current measured with the sample in the measurement position,
 ${I}_{\text{PMT,~s.~out}}$ is the PMT current measured with the sample out of the position.
Transparency measurements in ASSET was performed  in the VUV range, usually between 120 nm and 200 nm.
Transparency can also be assessed across the visible light spectrum, from 200 nm to 800 nm, using a~spectrophotometer (PerkinElmer, Lambda 650 UV/VIS Spectrometer \cite{SpectrometerTransparency}).

An important capability of the ASSET measurements is conducting ageing studies.
The procedure starts with an X-ray beam entering the irradiation chamber filled with a~gas mixture of Ar/CO$_2$ (70/30~\%) at ambient pressure that ionises the gas and creates primary electrons.
The primary electrons are attracted to multiplication wires, where an avalanche occurs in the high electric field region near the wires, producing additional electrons and ions.
The electrons from the avalanche are collected on the wires, while the ions move towards the grounded sample, where the charge is measured by the picoammeter \cite{Keithley}.
This process accumulates charge on the sample, allowing for the quantification of the drop in QE after exposure.
Further details about the ASSET photocathode characterisation device can be found in \cite{myMScThesis}.

\subsection{Characterisation with particle beams}

\begin{figure}[!t]
\begin{center}
\includegraphics[width=\columnwidth]{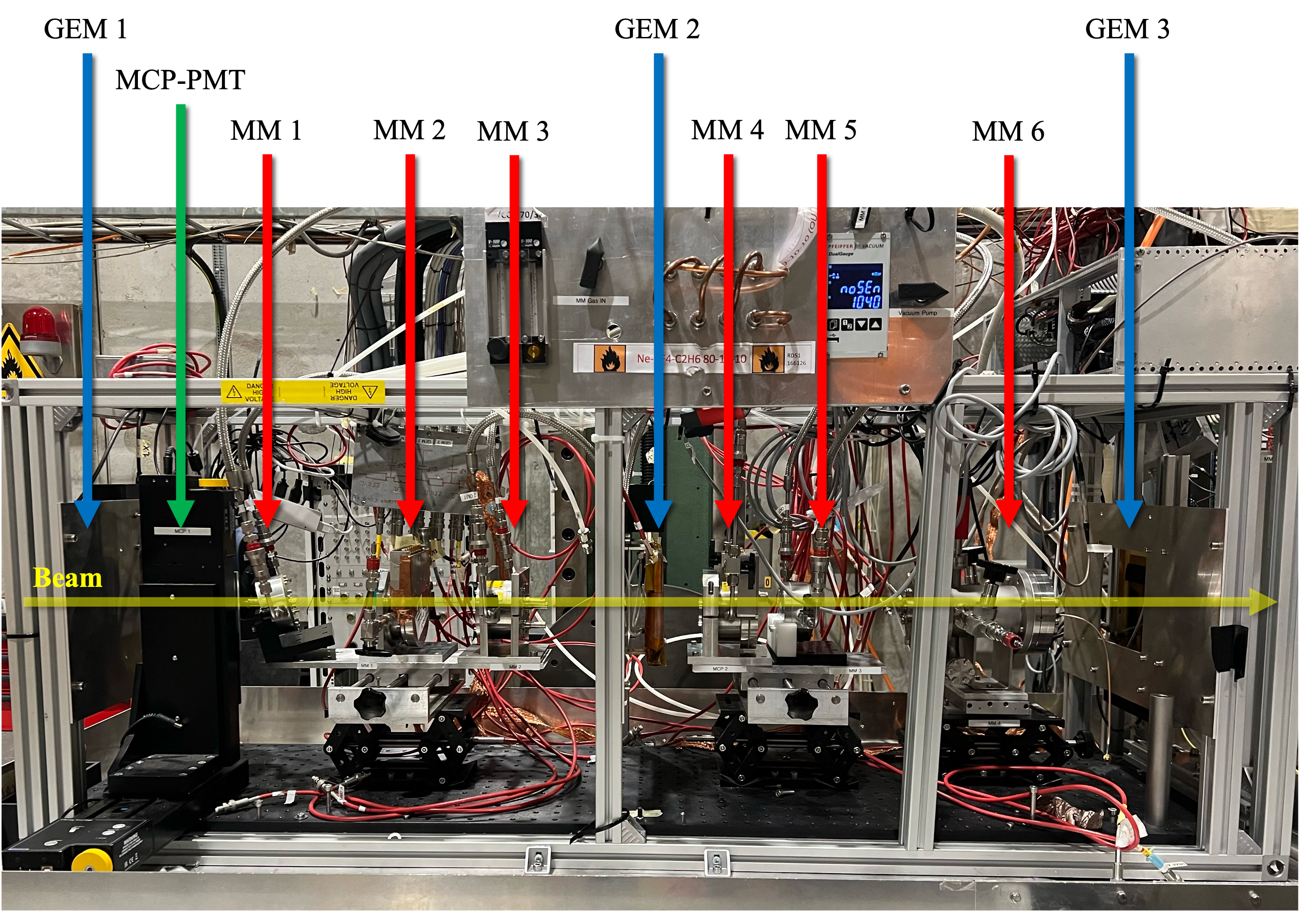}
\end{center}
\caption{The beam telescope equipped with three triple-GEMs for particle tracking, an MCP-PMT for timing reference as well as DAQ trigger, and PICOSEC prototypes for testing.}
\label{Telescope}
\end{figure}

Particle beam campaigns are conducted to measure the time resolution of prototypes assembled in various configurations.
The measurements are conducted using 150~GeV/c muon beams at the CERN SPS H4 beamline.
The test setup includes a~beam telescope facilitating triggering, timing and tracking capabilities.
An example telescope configuration is shown in Fig.~\ref{Telescope}.
Precise particle tracking is achieved utilising three triple Gas Electron Multiplier (GEM) detectors with a~spatial resolution below 80~µm.
These devices use APV25~\cite{APV25} front-end ASICs to shape the signals from all electrode channels, which are subsequently digitised using the Scalable Readout System (SRS) \cite{SRS}.
The GEMs are operated in a~gas mixture of Ar/CO$_2$~(70/30~\%) at ambient pressure.
A~micro-channel plate photomultiplier tube (MCP-PMT, Hamamatsu, R3809U-50~\cite{MCP-PMT}) serves as the timing reference and data acquisition (DAQ) trigger.
The telescope can be utilised for testing several PICOSEC prototypes simultaneously.

Different components individually contribute to the overall time resolution in a detector system, with the total time resolution $\sigma_{\text{tot}}$ being the sum of variances from each of these contributions.
It can be represented by a simplified model:
\begin{equation}
{\sigma_{\text{tot}}}^2= {\sigma_{\text{MIP}}}^2 +{\sigma_{\text{t}_0}}^2+{\sigma_{\text{e}}}^2+...
\label{eq:timeResolution}
\end{equation}
The first term, \(\sigma_{\text{MIP}}\), is the time resolution of the PICOSEC prototype measured with MIPs, including the detector's noise coming from amplitude fluctuations. There is a dependence between the time resolution and the number of photoelectrons:
\begin{equation}
\sigma_{\text{MIP}} = {\ {\sigma_{\text{SPE}}} \over {\sqrt{N_{\text{PE}}}} },
\label{eq:timeResolution2}
\end{equation}
where $\sigma_{\text{SPE}}$ is the time resolution achieved with single photoelecton (SPE) events and $N_{\text{PE}}$ is the number of photoelectrons extracted from a photocathode.
For photocathodes with higher $N_{\text{PE}}$, a better time resolution $\sigma_{\text{MIP}}$ is expected.
The second term, $\sigma_{\text{t}_0}$, in Equation~\ref{eq:timeResolution} is the contribution of the reference device.
The third term, $\sigma_{\text{e}}$, is the time jitter caused by the connected external electrical circuit.
The time resolutions discussed in this paper include contributions from all these components.

Quantifying the time resolution of the PICOSEC detector requires a~reference device with significantly superior timing precision.
To fulfill this condition, the MCP-PMT with a time resolution below 6 ps in the central region \cite{LukasPhDThesis} serves as the reference for the PICOSEC prototypes. 
The devices are aligned to each other, ensuring that particles passed through the active area of both.
To amplify and read out the signal, a~custom-developed RF pulse amplifier cards optimised for PICOSEC (with built-in discharge protection up to 350 V, 650 MHz bandwidth, 38 dB gain, 75 mW power consumption \cite{AntonijaMPGD, RFamp}) and an oscilloscope  (LeCroy WR8104, 10 GS/s sampling rate, 1~GHz bandwidth) are used.

In the analysis, the constant fraction discrimination  method is employed to determine the timestamp at which the signal reaches a~specific fraction of its peak amplitude, thereby addressing the issue of time walk \cite{firstPicosecPaper}.
To reduce noise interference in the timing determination, the leading edge of the signal is fitted using a~sigmoid function as it accurately represents its shape.
The signal position in time at 20$\%$ of its amplitude is identified.
The SAT values are computed as the difference between those timestamps, specifically by subtracting the MCP-PMT from the PICOSEC detector.
The histogram of SAT values is fitted with a double Gaussian function, which effectively approximates the shape of the distribution and accounts for the variations between the SAT and the e-peak charge.
The double Gaussian function is given by the equation:
\begin{equation}
f(\Delta t) = N \left[a~exp \left(-\frac{\Delta t - \mu}{2 \sigma^{2}_{\text{core}}}\right) + (1-a)~exp \left(-\frac{\Delta t - \mu}{2 \sigma^{2}_{\text{tail}}}\right) \right],
\label{eq:gauss}
\end{equation}
where \( N \) is a scaling factor,  \( a \in [0, 1] \) is a weighting parameter, \( \mu \) is the mean of the distribution, \( \sigma_{\text{core}} \) is the standard deviation of the core Gaussian and \( \sigma_{\text{tail}} \) is the standard deviation of the Gaussian describing the tail.
The weighting parameter between the two Gaussian components enables the determination of the combined standard deviation $\sigma_{\text{comb}}$ of the overall distribution:
\begin{equation}
\sigma^{2}_{\text{comb}} = a~\sigma^{2}_{\text{core}} + (1-a)~ \sigma^{2}_{\text{tail}}~.
\label{eq:gauss_comb}
\end{equation}
This combined standard deviation $\sigma_{\text{comb}}$ represents the time resolution of the detector system.
Finally, the efficiency of the detector equipped with a given photocathode is calculated by dividing the number of events that induce a signal on the PICOSEC prototype by the number of events that induce a signal on the MCP-PMT when a charged particle passed through both devices.

The results presented in this paper are derived from calculations conducted after applying several selection criteria (cuts) to the triggered events.
Initially, a~time window cut is implemented, selecting events within 300 ps of the median time difference of all recorded signals to exclude off-time events and noise fluctuations. 
Additional cuts are applied to set signal amplitude limits.
Events with amplitudes below~1\%~of the dynamic range, categorised as empty events, and those above~99\%, classified as saturated waveforms, are rejected.
Furthermore, a~geometrical cut is essential due to the Cherenkov radiator's emission of a~UV photon cone at approximately~45-degree angle.
To ensure accurate time resolution measurements, only fully contained events within the detector's active area should be included. 
Specifically, a~cut of a~4~mm diameter circle around the pad center for a~10~mm diameter detector equipped with a~3 mm thick radiator (or a~larger diameter, if otherwise specified) is implemented.
Signals from tracks that pass outside this central region show reduced amplitude because of the partial loss of photoelectrons beyond the area of the channel.
The efficiency of the detectors consistently remains above 95\%, despite the implementation of corrections and cuts to the triggered events.


The photocathodes performance is additionally assessed by comparing  $N_\text{PE}$  they generate during the process.
The number is estimated as a~ratio of the signal amplitude induced by an~SPE to the signal amplitude induced by a~MIP for a~detector at the same settings.
SPE measurements are performed using a~light-emitting diode (LED) to uniformly irradiate the photocathode and extract one photoelectron at a~time, while the multiple photoelectron measurements are conducted with a~muon beam.
For the SPE measurements, an additional measurement with the LED off was performed to record only the noise.
The signals are amplified by a~custom-developed charge-sensitive ARTH amplifier \cite{LukasPhDThesis} and read out by an oscilloscope.

\begin{figure*}[!t]
\begin{center}
\includegraphics[width=14cm]{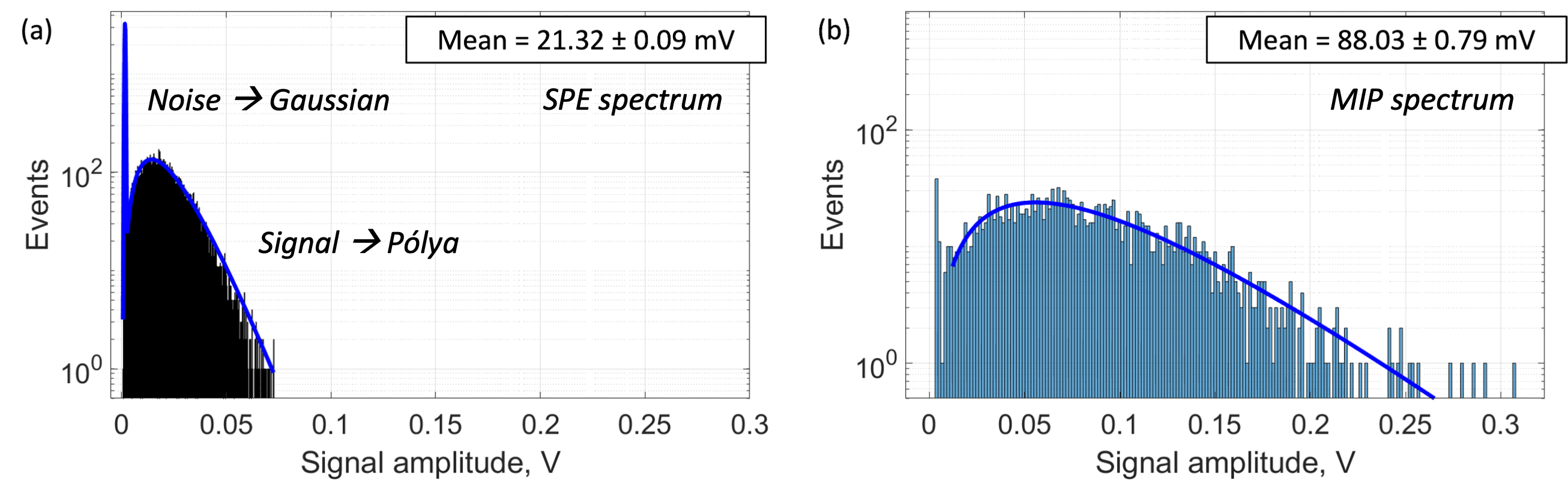}
\end{center}
\caption{Examples of histograms of the maximum amplitudes fitted with Gaussian and Pólya distributions for (a)~SPE and (b) MPE, with calculated mean values.}
\label{PolyaNPE}
\end{figure*}

The procedure for calculating  $N_\text{PE}$  involves several steps.
Firstly, the maximum amplitude for each waveform is determined.
Secondly, a~histogram of all maximum amplitudes is plotted.
For the LED-off measurement, the Gaussian distribution parameters are saved and later applied to the LED-on analysis.
In the LED-on measurement, the noise component is fitted with a Gaussian distribution, while the signal component is fitted using a Pólya distribution of the form:
\begin{equation}
P_n = \frac{(\theta + 1)^{\theta + 1}}{\bar{n} \Gamma (\theta + 1)}  \left( \frac{n}{\bar{n}} \right)^{\theta} e^{-(\theta + 1)n/\bar{n}},
\label{eq:Pólya}
\end{equation}
where: $n$ is the number of charges produced, $\bar{n}$ is the mean avalanche size and $\theta$ is the shape parameter.
In the muon beam measurement, only the Pólya distribution is used to fit the signal, since the noise component is negligible.
The mean amplitude for the MIP, $\bar{n}_{\text{MIP}}$, is divided by the mean amplitude for the SPE, $\bar{n}_{\text{SPE}}$, to obtain the $N_{\text{PE}}$ for a~given photocathode:
\begin{equation}
N_{\text{PE}}=\frac{ \bar{n}_{\text{MIP}} } { \bar{n}_{\text{SPE}}}.
\label{eq:NPE}
\end{equation}
Examples of histograms of the maximum amplitudes for SPE and MIP are presented in Fig. \ref{PolyaNPE}.

\section{Photocathode characterisation}
\label{sec:4}

Three different photocathode materials including Cesium Iodide, Diamond-Like Carbon and Boron Carbide have been examined.
The fabrication, characterisation and performance evaluation of these photocathode materials have been detailed, highlighting their strengths and challenges in achieving precise timing in advanced detection systems.
To ensure consistency of the results, all measurements presented in this paper were performed using identical single-pad non-resistive detectors featuring a~10~mm diameter active area (or 15 mm if otherwise specified), with pre-amplification and amplification gaps of approximately 127~µm.
The detectors were operated in a~sealed mode, maintaining a~gas pressure of 990~$\pm$~5~mbar for the purpose of comparison.

\subsection{Cesium Iodide}

Cesium Iodide is the most widely used material for photocathodes in gaseous radiation detectors due to its high QE and  sensitivity to UV radiation.
Notable applications of CsI as a~UV-to-electron converter include its use in the RICH detectors of the particle identification systems for ALICE \cite{ALICE_RICH, ALICE_CsI} and COMPASS~\cite{COMPASS_RICH_CsI}.
The baseline photocathode used in the PICOSEC detector consists of an 18 nm thick CsI layer deposited onto a~3 mm thick MgF$_2$ substrate with a~3 nm thick Cr conductive interfacial layer, exceeding 12 photoelectrons per MIP~\cite{firstPicosecPaper}.
Previous studies on transparency in the VUV range have shown that the pure MgF$_2$ substrate has approximately 80\% transparency, which decreases to 45\% with the addition of the Cr layer, and further to 10\% with the CsI layer \cite{myMScThesis}.

\begin{figure}[!b]
\begin{center}
\includegraphics[width=\columnwidth]{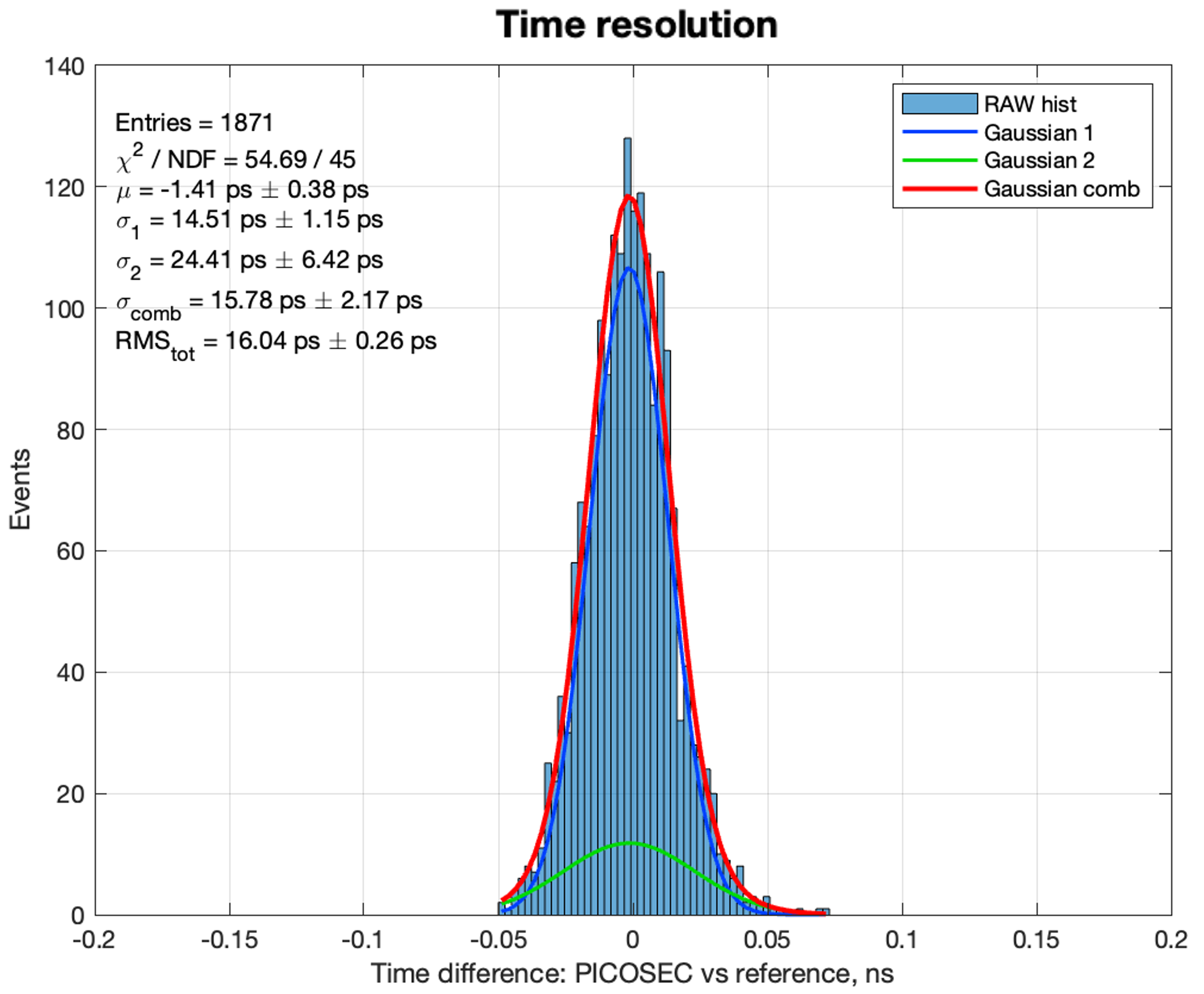}
\end{center}
\caption{SAT distribution of a~single-pad prototype, featuring a~10 mm diameter active area, equipped with an 18 nm thick CsI photocathode with a~3~nm thick Cr conductive layer, operated in a~sealed mode at a~gas pressure of 985~mbar. The voltage settings were V$_\text{C}$~=~-430~V on the cathode and  V$_\text{A}$~=~265~V on the anode. The histogram consists of the data after implementing a~geometrical cut of a~4 mm diameter circle around the pad center to include only fully contained events. The results of a~double Gaussian fit yield a~time resolution of $\sigma$~=~15.8~$\pm$~2.2~ps.}
\label{CsITimeRes}
\end{figure}

Particle beam measurements were performed using a PICOSEC prototype equipped with a CsI photocathode, operated in a sealed mode at a gas pressure of 985 mbar.
At voltage settings of V$_\text{C}$~=~-430~V on the cathode and  V$_\text{A}$~=~265~V on the anode, the detector exhibited a~time resolution of $\sigma$~=~15.8~$\pm$~2.2~ps,  as illustrated in Fig.~\ref{CsITimeRes}, with a full detection efficiency for completely contained events.

\begin{figure}[!t]
\begin{center}
\includegraphics[width=\columnwidth]{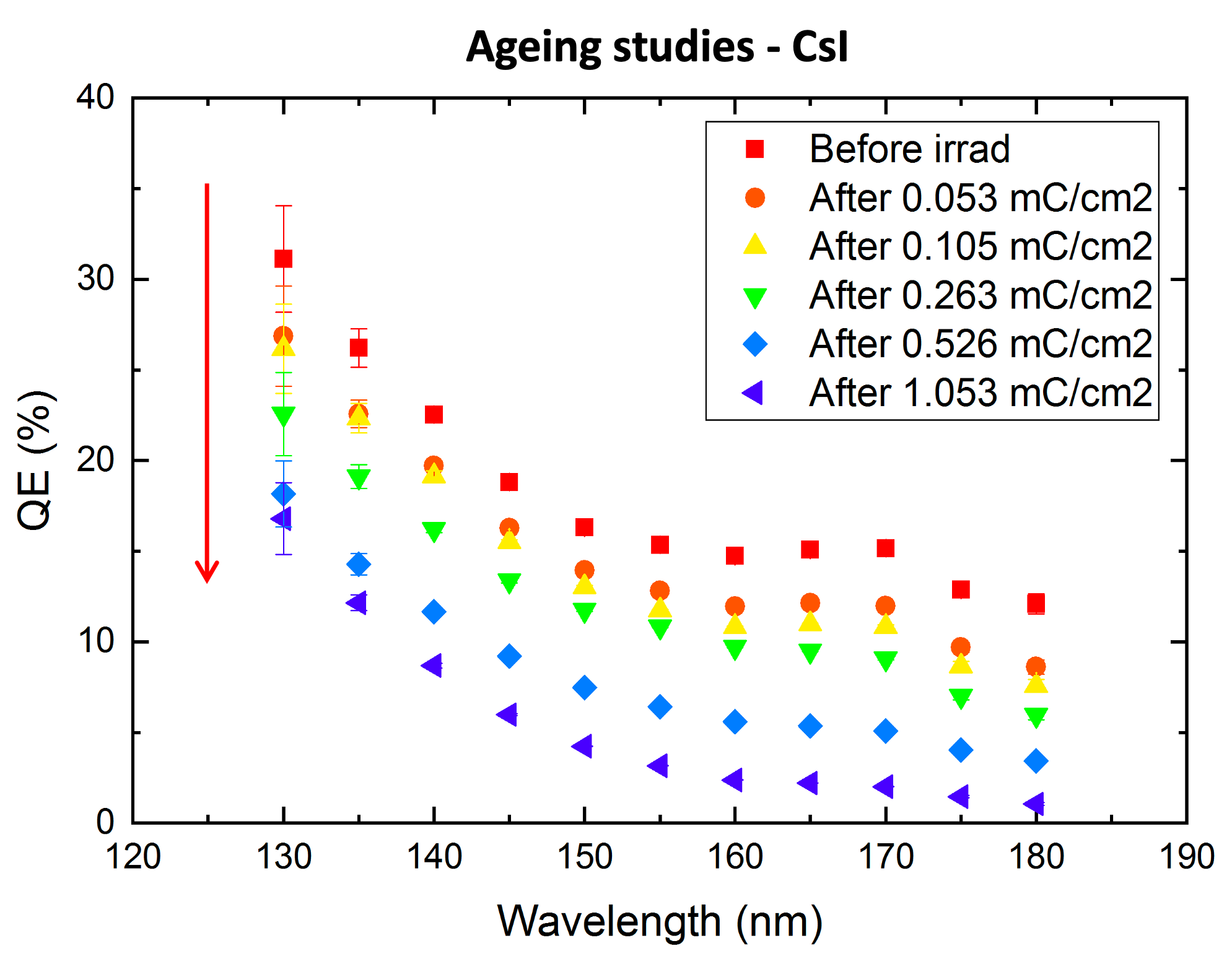}
\end{center}
\caption{Ageing studies of a~CsI photocathode performed in the ASSET setup. Decrease in QE after multiple irradiation steps indicates the impact of ion bombardment on CsI \cite{myMScThesis}.}
\label{CsIageing}
\end{figure}

Despite exhibiting excellent time resolution, CsI is vulnerable to damage from ion backflow and discharges.
Figure~\ref{CsIageing} illustrates the decline in QE observed in a~CsI sample before and after multiple irradiation steps, providing evidence of ion bombardment affecting CsI.
Studies demonstrated that after accumulating a~charge of 6 mC/cm$^2$, the QE for CsI decreased by 80\%, whereas for B$_4$C, it decreased by 45\%, indicating a~more rapid degradation of CsI \cite{myMScThesis}.
One possible solution to reduce the impact of ion backflow on CsI and mitigate degradation involves introducing protective layers, such as MgF$_2$ or Lithium Fluoride (LiF) \cite{PhotoProt}.
Nonetheless, tests conducted within the PICOSEC project revealed that these layers inhibited electron extraction, leading to reduced QE \cite{myMScThesis}.
Additionally, CsI photocathodes are sensitive to humidity and require storage in a~vacuum or dry gas to maintain good time resolution, as they degrade within minutes of humidity exposure. 
To address the issue of non-robustness, one approach is to explore alternative materials, with carbon-based photocathodes such as DLC and B$_4$C emerging as the most promising candidates.

\subsection{Diamond-Like Carbon}

Diamond-Like Carbon represents an alternative photocathode material that offers greater resistance to environmental influences compared to CsI.
DLC belongs to a~class of amorphous carbon materials, exhibiting excellent properties including chemical and thermal stability, hardness and robustness.
Initial measurements of DLC photocathodes have been conducted by collaborators from USTC  \cite{XuPhotocathodes}, including tests of various layer thicknesses, measurements of QE, $N_\text{PE}$ and time resolution as well as aging studies.
DLC samples consistently demonstrated good performance and robustness,  making them a~strong candidate for PICOSEC photocathodes.

The DLC photocathodes presented in this paper were fabricated using a~magnetron sputtering technique at the CERN Micro-Pattern Technologies (MPT) workshop.
DLC photocathodes with thicknesses ranging from 1 nm to 100 nm were deposited on glass and MgF$_2$ substrates, with some samples including a~Cr interfacial layer.
The process was reproducible across both campaigns.

\begin{figure}[!t]
\begin{center}
\includegraphics[width=\columnwidth]{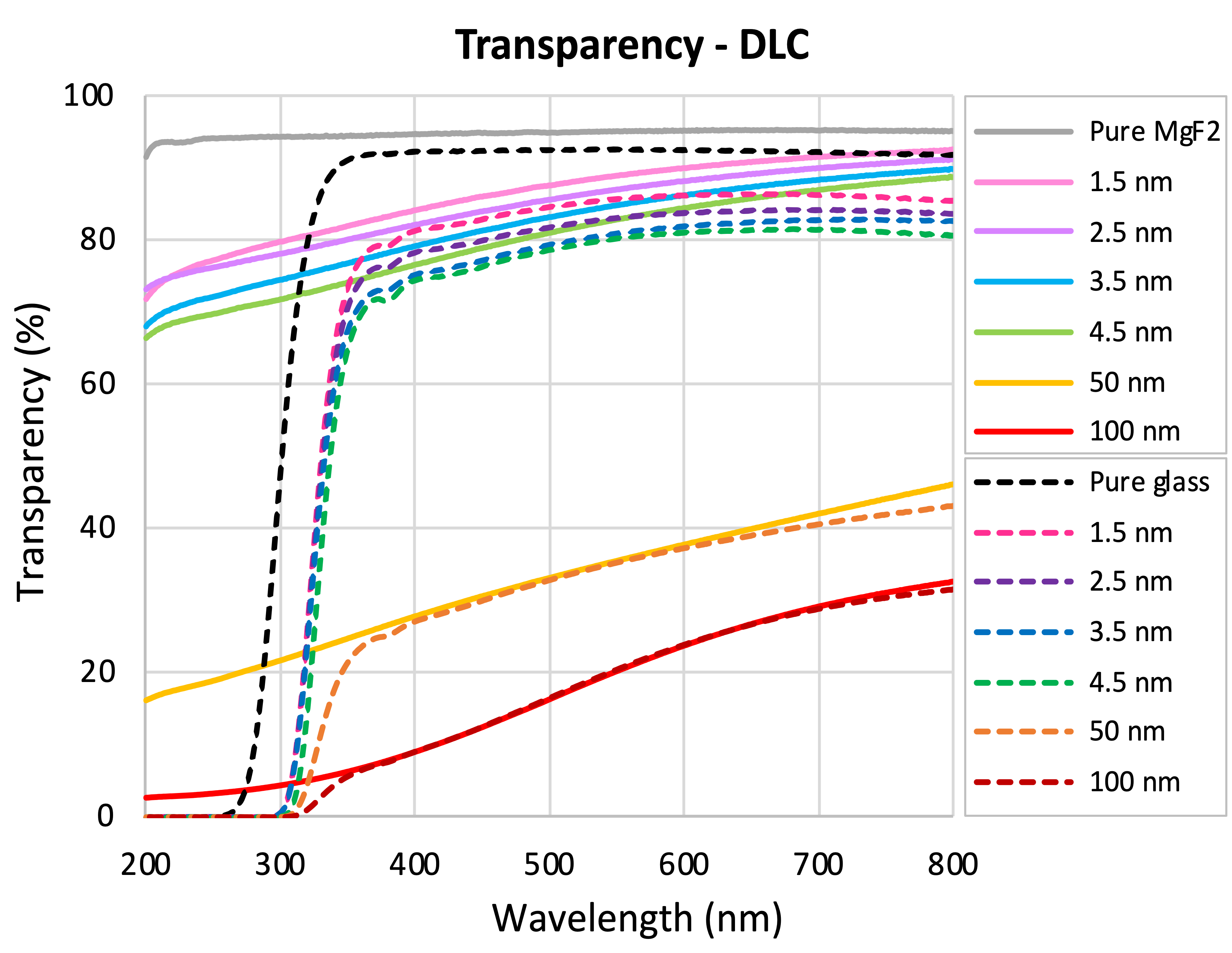}
\end{center}
\caption{Transparency measurements of the DLC layers with thicknesses ranging from 1.5 nm to 100 nm deposited on MgF$_2$ and glass substrates, conducted in the wavelength range from 200 nm to 800 nm using a~spectrophotometer.}
\label{DLCTransparency}
\end{figure}

\begin{figure}[!t]
\begin{center}
\includegraphics[width=\columnwidth]{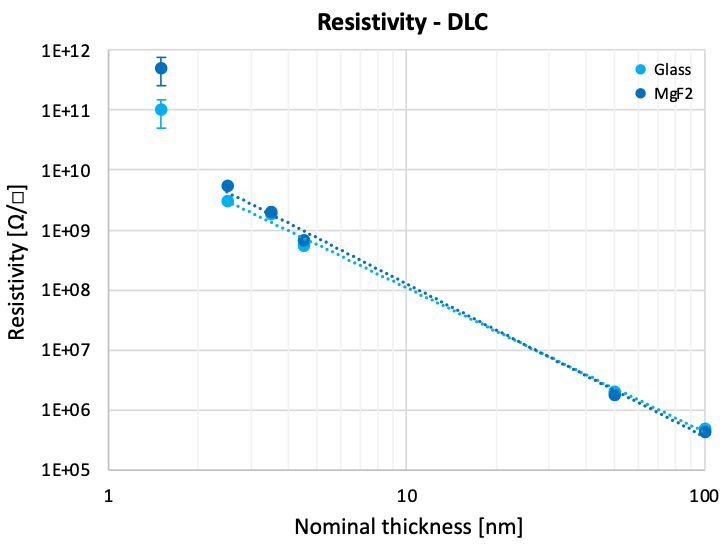}
\end{center}
\caption{Surface resistivity measurements of the DLC photocathodes, presenting a~correlation between  estimated thicknesses and resistivity. The higher resistivity of the DLC photocathodes on the MgF$_2$ radiator compared to the glass substrate suggests a~thinner deposited layer on the MgF$_2$, possibly due to lower crystal adhesion.}
\label{DLCResistivity}
\end{figure}

Profilometer measurements were performed to investigate the thicknesses of the 50~nm and 100~nm layers.
For layers in the range of a~few nanometers, specifically from 1.5~nm to 4.5~nm, thicknesses were estimated by scaling the coating time.
Figure~\ref{DLCTransparency} presents the results of transparency measurements conducted in the wavelength range from 200~nm to 800~nm using a~spectrophotometer.
The data demonstrate a~correlation between the estimated thicknesses and the measured transparency.
The DLC photocathodes deposited on both substrates exhibit similar behaviour; nevertheless, the glass samples show slightly reduced transparency and a low-wavelength cutoff below 300~nm.
The transparency of MgF$_2$ samples with a~thickness of a~few nanometers at 160 nm is about 60\%.

\begin{figure}[t]
\begin{center}
\includegraphics[width=\columnwidth]{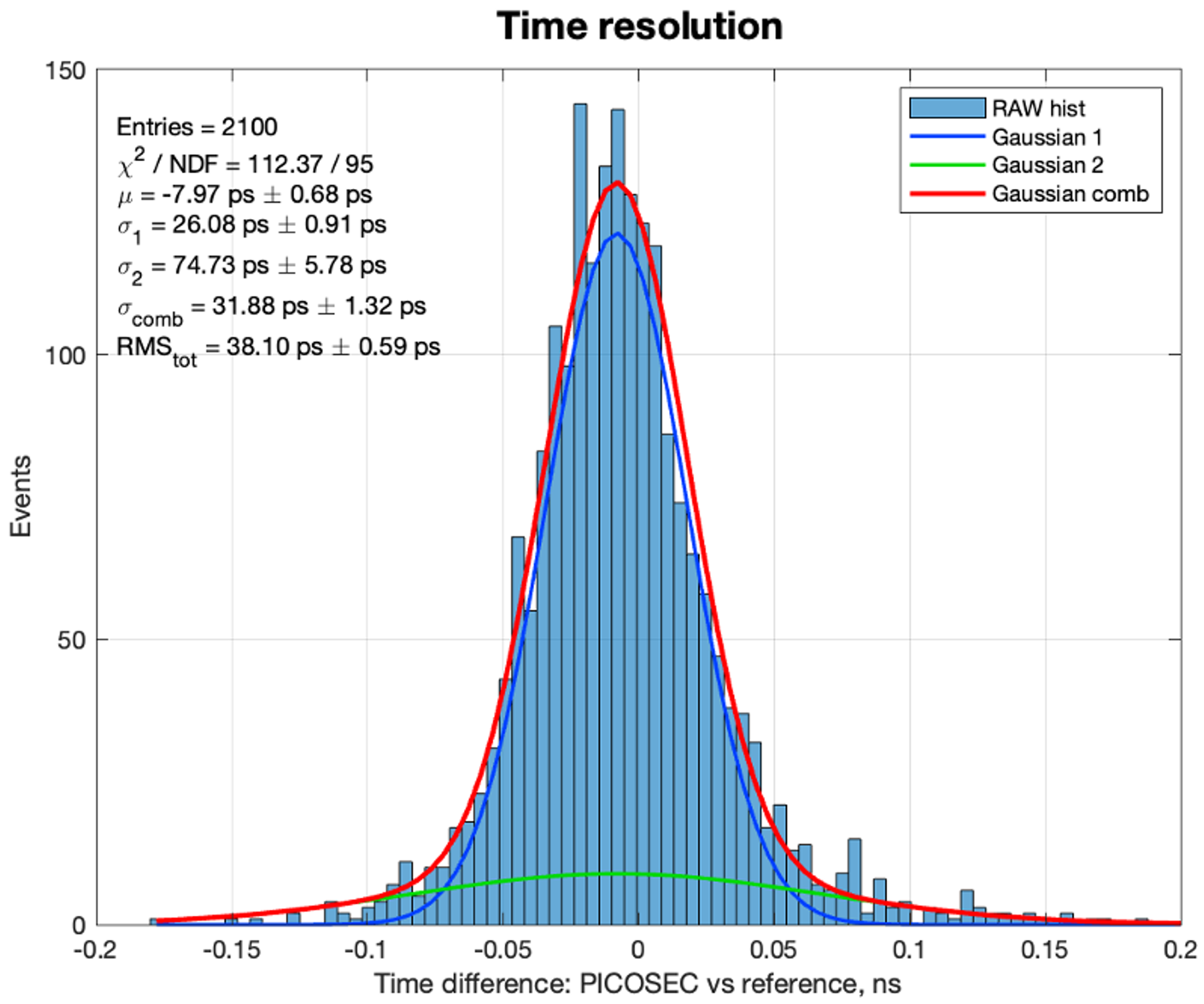}
\end{center}
\caption{SAT distribution of a~prototype equipped with a~1.5 nm DLC photocathode.
The voltage settings were  V$_\text{C}$~=~-500~V on the cathode and  V$_\text{A}$~=~275~V on the anode.
The results show a~time resolution of $\sigma$~=~31.9~$\pm$~1.3~ps.
The discrepancy in time resolution between $\sigma$ and RMS is attributed to the fact that RMS incorporates all data points, including outliers, which, while infrequent and relatively insignificant, nonetheless impact the overall timing performance.}
\label{DLCTimeRes}
\end{figure}

Surface resistivity measurements were conducted as an additional test to confirm whether the estimated thicknesses follow the expected trend, where thinner samples exhibit higher surface resistivity.
The measurements were performed using the~picoammeter \cite{Keithley}. A low voltage was applied across two conductive strips coated on the substrates and the resistivity of the photocathodes was recorded.
The results exhibit a~correlation between the thickness of the DLC layer and its resistivity, as shown in Fig.~\ref{DLCResistivity}.
The surface resistivity of the 1.5 nm DLC samples deviates from the expected trend, which may be attributed to the potential non-uniformity of these very thin layers.
Additionally, the higher values for the DLC photocathodes on the MgF$_2$ radiator compared to the glass substrate suggests that the layer deposited on the MgF$_2$ is thinner, potentially due to the lower adhesion of the crystal.

The DLC photocathodes with thicknesses ranging from 1.5~nm to 3.5 nm were characterised during particle beam measurements.
The measurements were performed at a~gas pressure of 990~mbar.
The $N_\text{PE}$ generated by the DLC photocathodes varied between 2.5 and 3.
The 10 mm diameter active area detector featuring a~1.5 nm thick DLC photocathode deposited directly on the radiator and operated at voltages of  V$_\text{C}$~=~-500~V on the cathode and  V$_\text{A}$~=~275~V on the anode demonstrated the best performance, achieving a~time resolution of $\sigma$~=~31.9~$\pm$~1.3~ps, as illustrated in Fig.~\ref{DLCTimeRes}, with a detection efficiency of 96.8\%.
Thicker samples exhibited comparable time resolution values, with approximately 4~ps worse performance.

The DLC photocathodes deposited directly on the radiator without a~Cr conductive interfacial layer are sufficient for studying photocathode performance.
Nonetheless, in high-rate environments, a~Cr layer is essential to mitigate charging-up effects and voltage drops, particularly for samples with larger surface areas.
Consequently, samples with Cr were evaluated, showing a~decrease in transparency of approximately 30\% and 2 ps worse time resolution.

To enhance UV photon production, a~thicker radiator was introduced.
At the same time, the Cherenkov cone diameter was widened.
A prototype with a~15 mm diameter active area, equipped with a~2.5 nm thick DLC photocathode deposited on a~5 mm thick MgF$_2$ radiator, was tested.
The analysis involved applying a~geometric cut, consisting of a~5 mm diameter circle around the pad center, to include only fully contained events.
The detector operated at of  V$_\text{C}$~=~-490~V on the cathode and  V$_\text{A}$~=~275~V on the anode yielded a~time resolution of $\sigma$~=~28.0~$\pm$~1.4 ps, with a 99.4\% detection efficiency.

\begin{figure}[!t]
\begin{center}
\includegraphics[width=\columnwidth]{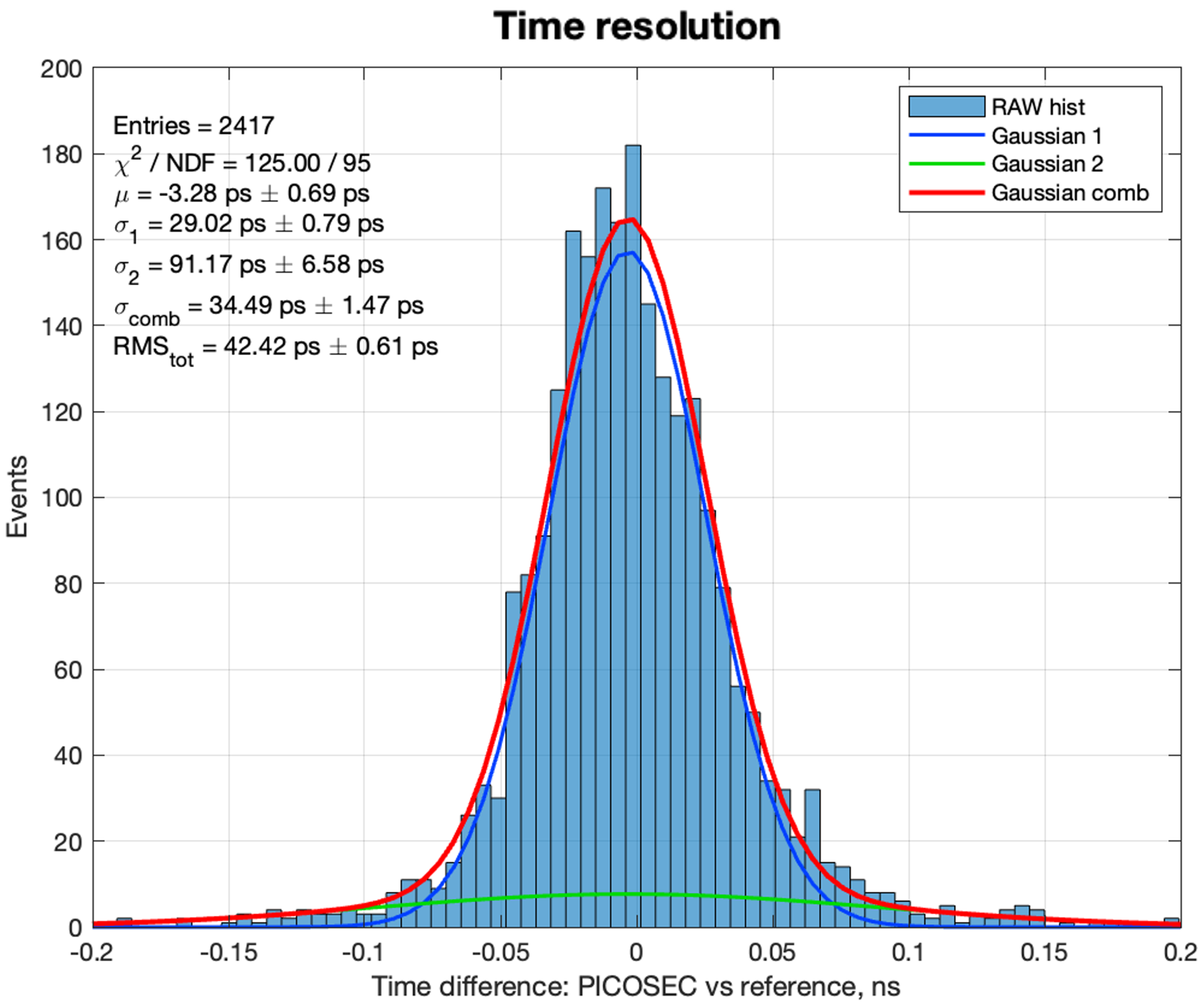}
\end{center}
\caption{SAT distribution of a~prototype equipped with a~9 nm B$_4$C photocathode.
The voltages were at V$_\text{C}$~=~-490~V on the cathode and  V$_\text{A}$~=~275~V on the anode.
The results show a~time resolution of $\sigma$~=~34.5~$\pm$~1.5~ps.}
\label{B4CTimeRes}
\end{figure}

\subsection{Boron Carbide}

Boron Carbide is a~chemical compound from the carbide group, known for its crystalline structure and exceptional hardness, often used as a~substitute for diamond.
The initial B$_4$C photocathodes were developed at CEA, where a~wide range of thicknesses was tested, yielding promising results \cite{LukasPhDThesis,myMPGD}.
More detailed studies discussed in this paper were conducted with B$_4$C samples fabricated using a~sputtering technique at ESS.
During the coating campaigns, B$_4$C photocathodes with thicknesses ranging from 7~nm to 15~nm were deposited on MgF$_2$ radiators with a~Cr conductive interfacial layers.

Profilometer measurements were conducted to determine the thickness of the photocathodes.
The 7 nm layer was estimated based on the scaling of the coating time, considering the machine's resolution limits.
Transparency measurements of the B$_4$C layers, conducted in the VUV wavelength range, demonstrated a~correlation between estimated thicknesses and transparency, with values ranging from 40\% for the thinnest layer to 20\% for the thickest.

The B$_4$C photocathodes were evaluated through particle beam measurements.
The prototypes were operated at a~gas pressure of 990 mbar.
The $N_\text{PE}$ created by the B$_4$C photocathodes ranged between 2 and 4, with higher photoelectron production observed for thinner layers.
The detector equipped with a~9~nm thick B$_4$C photocathode and operated at  V$_\text{C}$~=~-490~V on the cathode and  V$_\text{A}$~=~275~V on the anode, exhibited the best performance, achieving a~time resolution of $\sigma$ = 34.5~$\pm$~1.5~ps, as shown in Fig.~\ref{B4CTimeRes}, with a 95.4\% detection efficiency.
Thicker B$_4$C samples showed worse time resolutions, with the thickest sample by around 10 ps.

\section{Discussion}
\label{sec:5}

In the comparative analysis of photocathode materials, distinct performance characteristics were observed among CsI, DLC, and B$_4$C.
CsI exhibited the highest $N_\text{PE}$, exceeding 12~photoelectrons per MIP, whereas DLC yielded $N_\text{PE}$ ranging between 2.5 and 3, while B$_4$C between 2 and 4.
Despite B$_4$C being more transparent than DLC, both materials exhibit similar efficiency, generating comparable $N_\text{PE}$, which correlates with the achievable time resolutions.
CsI showed the best performance with $\sigma$~$\approx$~15.8~ps, followed by DLC with $\sigma$~$\approx$~32~ps and B$_4$C with $\sigma$ $\approx$ 34.5 ps.
Although CsI achieved excellent time resolution, it showed faster degradation in QE due to ion backflow, compared to carbon-based materials.
Additionally, the carbon-based photocathodes are not sensitive to humidity and have been successfully stored in air without any visual or performance degradation.
These samples have been reused across multiple test beam campaigns over several years, demonstrating no deterioration in time resolution.

\begin{figure}[!t]
\begin{center}
\includegraphics[width=\columnwidth]{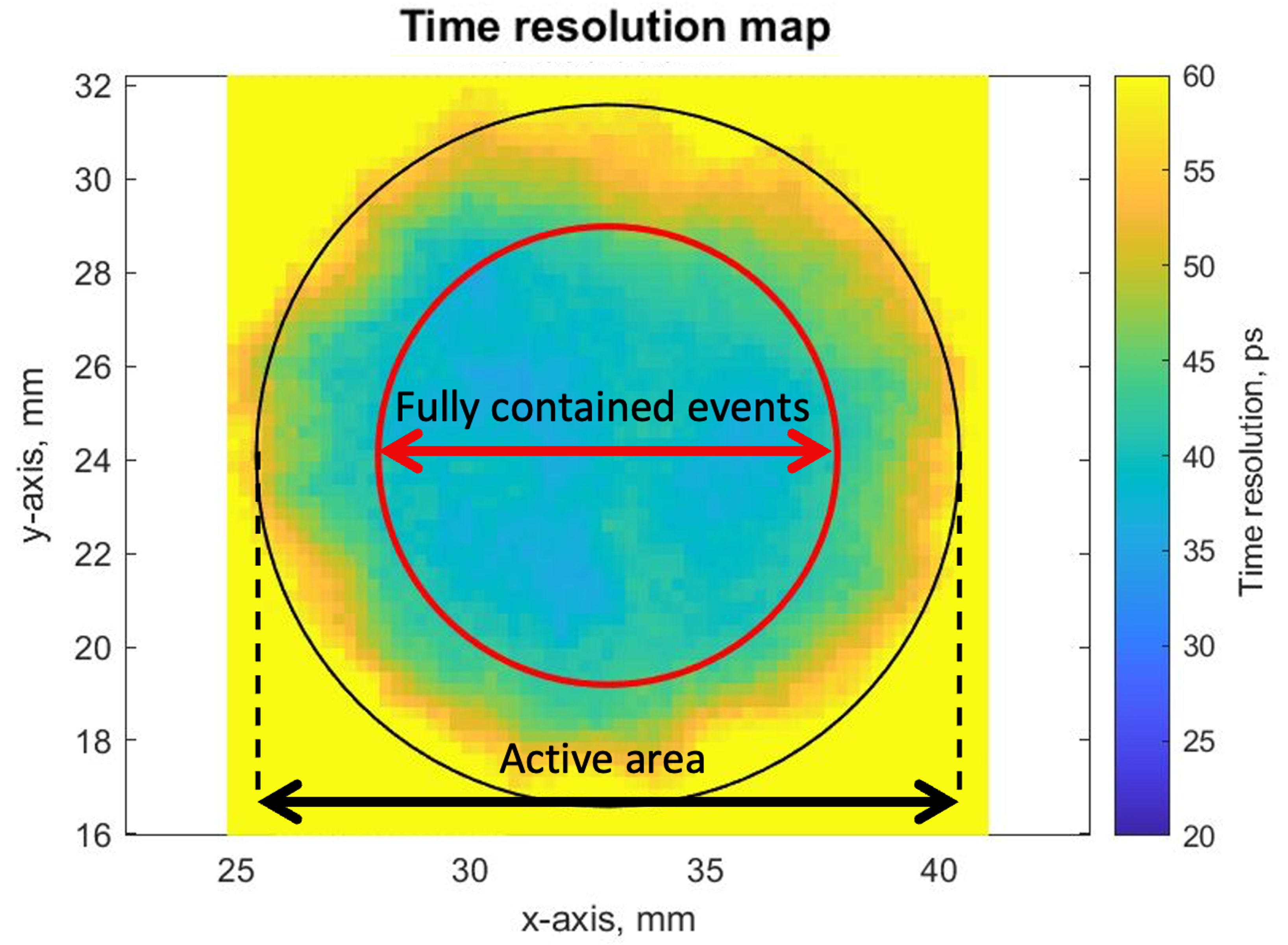}
\end{center}
\caption{Time resolution map of a~detector configuration consisting of a~15~mm diameter active area resistive Micromegas of 20 M$\Omega$/$\Box$, a~1.5 nm DLC photocathode and an integrated amplifier. The black circle indicates the active area of the detector, while the red circle highlights fully contained events. The variations in time resolution observed in the center of the active area may suggest an issue with the photocathode's inhomogeneity.}
\label{RobustDetector}
\end{figure}

Considering the demand for robust detectors with timing specifications that allow a~margin on operating condition, DLC and B$_4$C emerge as promising candidates.
Consequently, a~detector configuration comprising a~single-pad prototype with a~15~mm diameter active area resistive Micromegas of 20~M$\Omega$/$\Box$, utilising a~1.5~nm DLC photocathode and an amplifier integrated on the outer PCB, was tested.
Operating the device at  V$_\text{C}$~=~-530~V on the cathode and  V$_\text{A}$~=~275~V on the anode, a~time resolution of $\sigma$~=~31.4 $\pm$ 0.6 ps was achieved within a~9~mm diameter circle around the pad center, including only fully contained events.
A good time response across this region, with an RMS~=~38.8~$\pm$~0.3 ps, was observed, as depicted in Fig.~\ref{RobustDetector}. A~detection efficiency was around 97.6\%.

\section{Conclusions}
\label{sec:6}

The work described in this paper focuses on enhancing the robustness of PICOSEC detectors.
The research delves into the characterisation of carbon-based photocathodes, such as DLC and B$_4$C.
Results obtained from prototypes equipped with these photocathodes showed time resolutions of approximately $\sigma$~$\approx$~32~ps and $\sigma$~$\approx$~34.5~ps, respectively.
Furthermore, carbon-based materials displayed notably greater resistance to ion backflow, discharges and humidity. 
The efforts dedicated to advancing the detectors increase the feasibility of the PICOSEC concept for experiments requiring sustained performance while maintaining excellent timing precision.

Current PICOSEC developments include research on alternative materials, such as titanium to replace Cr, and carbon-based nanostructures for use as photocathodes.
Regarding the robustness of photocathodes under different humidity conditions, specific measurements are planned.
Simultaneously, a~10×10~cm$^2$ robust photocathode, incorporating a~conductive interlayer to prevent a~voltage drop, will be tested with a~100-channel prototype \cite{NDIP,AntonijaMPGD} and a~SAMPIC digitiser \cite{SAMPIC}.
In view of improving the stability of the detector for high-rate conditions, the production of a~10×10 cm$^2$ Micromegas with double-layer DLC for vertical charge evacuation and evaluation of rate capability is ongoing.
Additionally, efforts to enhance spatial resolution involve testing high-granularity layouts.
Scaling up the PICOSEC detector by tiling 10×10~cm$^2$ modules or developing larger prototypes are the next steps.

\section*{Acknowledgements}
\label{sec:8}

We acknowledge the  support of the CERN EP R\&D Strategic Programme on Technologies for Future Experiments; the RD51 Collaboration, in the framework of RD51 common projects; the DRD1 Collaboration; the PHENIICS Doctoral School Program of Université Paris-Saclay, France; the Cross-Disciplinary Program on Instrumentation and Detection of CEA, the French Alternative Energies and Atomic Energy Commission; the French National Research Agency (ANR), project id ANR-21-CE31-0027; the Program of National Natural Science Foundation of China, grants number 11935014 and 12125505; the COFUND-FP-CERN-2014 program, grant number 665779; the Fundação para a~Ciência e~a Tecnologia (FCT), Portugal; the Enhanced Eurotalents program, PCOFUND-GA-2013-600382; the US CMS program under DOE contract No. DE-AC02-07CH11359; this material is based upon work supported by the U.S. Department of Energy, Office of Science, Office of Nuclear Physics under contracts DE-AC05-06OR23177.


\bibliographystyle{elsarticle-num-names} 


\end{document}